\documentstyle[12pt]{article}

\makeatletter
\def\eqnarray{\stepcounter{equation}\let\@currentlabel=\theequation
\global\@eqnswtrue
\global\@eqcnt\z@\tabskip\@centering\let\\=\@eqncr
$$\halign to \displaywidth\bgroup\@eqnsel\hskip\@centering
  $\displaystyle\tabskip\z@{##}$&\global\@eqcnt\@ne
  \hfil$\displaystyle{\hbox{}##\hbox{}}$\hfil
  &\global\@eqcnt\tw@ $\displaystyle\tabskip\z@
  {##}$\hfil\tabskip\@centering&\llap{##}\tabskip\z@\cr}
\@addtoreset{equation}{section}
  \def\theequation{\thesection.\arabic{equation}}
\makeatother

\advance \textheight by \topskip
\begin{document}

\thispagestyle{empty}
\begin{flushright}
{\bf NF/DF-04/96}\\
{\bf hep-th/9701065}
\end{flushright}
$\ $\vskip 2truecm
\begin{center}

{ \Large \bf $R$-deformed Heisenberg algebra}\\
\vskip1.0cm
{ \bf Mikhail S. Plyushchay{}\footnote{E-mail: mikhail@fisica.ufjf.br 
and plushchay@mx.ihep.su}}\\[0.3cm]
{\it Departamento de F\' {i}sica -- ICE}\\
{\it Universidade Federal de Juiz de Fora}\\
{\it 36036-330 Juiz de Fora, MG Brazil}\\
{\it and}\\
{\it Institute for High Energy Physics}\\
{\it Protvino, Moscow Region, 142284 Russia} 
\end{center}
\vskip2.0cm
\begin{center}
{\bf Abstract}
\end{center}
It is shown that the deformed Heisenberg algebra involving the
reflection operator $R$ ($R$-deformed Heisenberg algebra) has
finite-dimensional representations which are equivalent to
representations of paragrassmann algebra with a special
differentiation operator.  Guon-like form of the algebra,
related to the generalized statistics, is found.  Some
applications of revealed representations of the $R$-deformed
Heisenberg algebra are discussed in the context of OSp(2$|$2)
supersymmetry.  It is shown that these representations can be
employed for realizing (2+1)-dimensional supersymmetry.  They
give also a possibility  to construct a universal spinor set of
linear differential equations describing either fractional spin
fields (anyons) or ordinary integer and half-integer spin fields
in 2+1 dimensions.
\vskip1.0cm
\begin{center}
{\it Mod. Phys. Lett.} {\bf A11} (1996) 2953-2964
\end{center}
\newpage

\section{Introduction}
The deformed Heisenberg algebra involving the reflection
operator $R$ has found many interesting physical applications.
It appeared naturally in the context of parafields
\cite{ok,mac1}, but earlier it was known
in connection with some quantum mechanical systems \cite{yang}.
Recently this algebra was used for investigating quantum
mechanical $N$-body Calogero model \cite{poly}, for bosonization
of supersymmetric quantum mechanics \cite{bem,bos,bosany} and
describing  anyons in (2+1) \cite{bosany,any} and (1+1)
dimensions \cite{any1}.  In all the listed applications the
infinite-dimensional unitary representations of the $R$-deformed
Heisenberg algebra were used.

In the present paper it will be shown that this algebra
has also finite dimensional representations which are equivalent to
representations of some paragrassmann algebra \cite{fik} with 
differentiation operator realized in a special form.
We shall show that guon-like algebra \cite{guon} can be
constructed in a natural way proceeding from the $R$-deformed
Heisenberg algebra.
Such guon-like algebra can be related in some way to the
$q$-deformed Heisenberg algebra \cite{qdef,macbi} with
deformation parameter $q$ being a primitive root of unity
\cite{fik}.  
We shall discuss some applications of finite-dimensional
representations of the $R$-deformed Heisenberg algebra.  In
particular, they will be used for realization of OSp(2$|$2)
supersymmetry.  The relationship of revealed representations to
finite-dimensional representations of $(2+1)$-dimensional
Lorentz group  will be established.  The latter will be employed
for realizing (2+1)-dimensional supersymmetry.  We shall also
use them for constructing a universal spinor set of linear
differential equations describing either fractional spin fields
(anyons) or ordinary integer or half-integer spin fields in 2+1
dimensions.
 
\section{Representations of $R$-deformed Heisenberg algebra}
The $R$-deformed Heisenberg algebra is given by the generators 
$a^-$, $a^+$, $1$ and by the reflection operator $R$
satisfying the (anti)commutation relations \cite{ok}--\cite{any}:
\begin{equation}
[a^-,a^+]=1+\nu R,\quad
R^2=1,\quad 
\{a^\pm,R\}=0,
\label{rdef}
\end{equation}
and $[a^\pm,1]=[R,1]=0$,
where $\nu\in {\bf R}$ is a deformation parameter.
The reflection operator $R$ is hermitian, 
whereas $a^+$ and $a^-$ will be considered as 
mutually conjugate operators with respect to 
appropriate scalar product.
One introduces the vacuum state $|0\rangle $,
$a^-|0\rangle =0$, $\langle 0|0\rangle =1$, $R|0\rangle =|0\rangle $,
and defines the 
states $|n\rangle =C_n (a^+)^n|0\rangle$
with some normalization constants $C_n$.
Then, from the relation
\begin{equation}
[a^-,(a^+)^n]=
\left(n+\frac{1}{2}(1-(-1)^n)\nu R\right)(a^+)^{n-1}
\label{rn}
\end{equation}
one concludes that algebra (\ref{rdef})
has infinite-dimensional unitary representations when 
$
\nu>-1. 
$
In this case the states $|n\rangle $ with 
$C_n=([n]_\nu !)^{-1/2}$,
$[n]_\nu !=\prod_{l=1}^n [l]_\nu$, $[l]_\nu=l+\frac{1}{2}(1-(-1)^l)\nu$,
form the complete orthonormal basis of Fock space representation,
$\langle n|n'\rangle =\delta_{nn'}$.
The reflection operator can be realized in terms of 
creation and annihilation operators via the number 
operator \cite{mac1,bosany},
\begin{equation} 
N=\frac{1}{2}\{a^+,a^-\}-\frac{1}{2}(\nu+1),\quad
N|n\rangle =n|n\rangle ,
\label{Nn}
\end{equation}
\begin{equation}
R=(-1)^N=\cos\pi N.
\label{RN}
\end{equation}
On the other hand, one can consider $R$-deformed 
Heisenberg algebra
(\ref{rdef}) working in the 
Schr\"{o}dinger 
representation, $\Psi=\Psi(x)$, with
creation-annihilation operators realized in the 
usual form 
$a^\pm=\frac{1}{\sqrt{2}}(x\mp ip)$.
Here the deformed momentum operator is \cite{ok}
$
p=-i(\frac{d}{dx}-\frac{\nu}{2x}R),
$
and operator $R$ acts as 
$R\Psi(x)=\Psi(-x)$, and so, 
$R\Psi_\pm(x)=\pm\Psi_\pm(x)$, 
$\Psi_\pm(x)=\Psi(x)\pm\Psi(-x)$. 
This explains the name of operator $R$.
One can note that if we write realization (\ref{RN})
in the Schr\"{o}dinger representation just in the case of
non-deformed ($\nu=0$)
Heisenberg algebra, we shall reveal
a hidden nonlocal nature of the reflection operator,
$R=\sin H_0$,
$H_0=\frac{1}{2}(x^2-d^2/dx^2)$.
Therefore, the reflection operator has a nature
similar to the nonlocal nature of the Klein operator \cite{ok}.

One can get the realization of the $R$-deformed Heisenberg algebra
in terms of non-deformed algebra with creation-annihilation
operators $b^\pm$ obeying the commutation relation
$[b^-,b^+]=1$. For the purpose, one represents
the operators $a^\pm$ as 
$a^-=F(N_b)b^-$, $a^+=(a^-)^\dagger=b^+F(N_b)$
with $F=F^\dagger$ being a function of the 
number operator $N_b=b^+b^-$.
Let us substitute these expressions for $a^+$ and $a^-$ and
$R=(-1)^{N_b}$ into the first relation from (\ref{rdef}),
and act on the complete set of orthonormal 
states  $|n\rangle_b\equiv(n!)^{-1/2}(b^+)^n|0\rangle =|n\rangle$, 
$N_b|n\rangle_b=n|n\rangle_b$, where
$|n\rangle$, $n=0,1,\ldots,$
are the Fock space states of deformed algebra.
As a result, we arrive  
at the sought for realization of deformed creation-annihilation
operators in terms of non-deformed ones,
\begin{equation}
a^-=F(N_b)b^-,\quad a^+=b^+F(N_b),
\label{abba}
\end{equation}
\begin{equation}
F(N_b)=\sqrt{1+\frac{\nu}{2(N_b+1)}
\left(1+(-1)^{N_b}\right)},\quad
\nu>-1.
\label{ab}
\end{equation}
Function (\ref{ab}) takes zero values if we put $\nu=-(2p+1)$,
$p=1,2,\ldots$. This indicates \cite{mmp}
that at these special values of the deformation parameter
algebra (\ref{rdef}) has finite-dimensional representations.
Then, using eq. (\ref{rn}), one finds that
for
$
\nu=-(2p+1),\quad
p=1,2,\ldots,
$
the relation
$
\langle\langle m|n\rangle\rangle=0,
$
$|n\rangle\rangle\equiv(a^+)^n|0\rangle$,
takes place for $n\geq 2p+1$ and arbitrary $m$. 
This means, in turn, that the relations $(a^+)^{2p+1}=(a^-)^{2p+1}=0$
are valid in this case. These latter relations specify
finite-dimensional representations of the   
$R$-deformed Heisenberg algebra.
Since in such representations for any
$p=1,2,\ldots$, there are the states with negative norm
(see eq. (\ref{rn})),
it means that these finite-dimensional Fock space
representations are non-unitary.

\section{$R$-paragrassmann algebra}
Let us consider the revealed finite-dimensional representations
in more detail.
We have arrived at the nilpotent algebra
\begin{eqnarray}
&&[a^-,a^+]=1-(2p+1)R,\quad
(a^\pm)^{2p+1}=0,\quad
p=1,2,\ldots,\label{aaa}\\
&&\{a^\pm,R\}=0,\quad
R^2=1.
\label{parag}
\end{eqnarray}
One can interpret $a^+$ as a
paragrassmann variable $\theta$, $\theta^{2p+1}=0$,
and in this case $a^-$ can be considered as
a differentiation operator \cite{fik}. Therefore,
the algebra (\ref{aaa}), (\ref{parag}) 
is a paragrassmann algebra of 
order $2p+1$ \cite{fik} with a special differentiation operator
whose action can be defined by relation (\ref{rn}).
We shall call it the R-paragrassmann algebra. 
Here, in addition to universal representation (\ref{Nn}),
(\ref{RN}),
one has also the normal ordered representation for the operator $R$,
\begin{equation}
R=\sum_{n=0}^{2p}f_{n}a^{+n}a^{-n},
\label{normr}
\end{equation}
with finite recursive relations defining coefficients 
$f_n$, 
\[
2f_{n-1}+[n]_\nu f_n-(2p+1)\sum_{i=0}^{[n/2]-1}f_{2i+1}
f_{n-(2i+1)}=0,\quad
n=1,\ldots, 2p,
\]
where $f_0=1$ and $[n/2]$ is an integer part
of $n/2$.

As a consequence of eqs. (\ref{aaa}), (\ref{parag}),
we have the relations
$
(1-R)a^{+2p}=(1-R)a^{-2p}=0.
$
They are equivalent to the nilpotency conditions 
$a^{\pm(2p+1)}=0$.
Besides, here operators $a^\pm$ satisfy the relation
\begin{equation}
a^+a^{-2p}+a^-a^+a^{-(2p-1)}+\ldots 
a^{-(2p-1)}a^+a^-+a^{-2p}a^+=0
\label{parasu}
\end{equation}
and corresponding conjugate relation.  Relations of form
(\ref{parasu}) take place in parasupersymmetric quantum
mechanics \cite{rubspi}.

As it has been mentioned above, finite-dimensional
Fock space representations of $R$-deformed Heisenberg algebra (\ref{rdef})
contain the states with negative norm. 
One may introduce the normalized states as
$|n\rangle=|\langle\langle n|n\rangle\rangle|^{-1/2}|n\rangle\rangle$.
They define the metric operator $\eta=\eta^\dagger$, $\eta^2=1$, 
whose matrix elements are
$||\eta||_{mn}=||\langle m|n\rangle||=diag(1,-1,-1,+1,+1,-1,-1,\ldots,
(-1)^{p-1},(-1)^{p-1},(-1)^p,(-1)^p)$.
With this metric operator, the indefinite
scalar product is given by the relation 
$(\Psi_1,\Psi_2)=\langle\Psi_1|\eta\Psi_2\rangle=
\Psi^*_{1n}\eta_{nm}\Psi_{2m}$, where $\Psi_n=\langle n|\Psi\rangle$.
The operators $a^+$ and $a^-$ can be
represented by the matrices
$
(a^+)_{mn}=A_n\delta_{m-1,n},$
$
(a^-)_{mn}=B_m\delta_{m+1,n},
$
with
$
A_{2k+1}=-B_{2k+1}=\sqrt{2(p-k)},$
$k=0,1,\ldots,p-1,$
$A_{2k}=B_{2k}=\sqrt{2k},$
$k=1,\ldots,p.$
They satisfy the relation
$(a^-)^\dagger=\eta a^+\eta$, and, as a consequence,
are mutually conjugate operators with respect
to this scalar product, $(\Psi_1,a^-\Psi_2)^*=
(\Psi_2,a^+\Psi_1)$.
The reflection operator has here the diagonal form
$R=diag(+1,-1,+1,\ldots,-1,+1)$.

Below we shall reveal the `physical explanation'
of non-unitarity of finite-dimensional
representations of the $R$-deformed Heisenberg
algebra in which $a^+$ and $a^-$ are 
interpreted as mutually conjugate operators.
On the other hand, one can define
hermitian conjugate operators $f^+=a^+$, $f^-=a^-R$,
in terms of which $R$-paragrassmann algebra 
(\ref{aaa}), (\ref{parag}) is rewritten equivalently as
\begin{equation}
\{f^+,f^-\}=(2p+1)-R,\quad
\{R,f^\pm\}=0,\quad
R^2=1,\quad
(f^\pm)^{2p+1}=0,\quad
p=1,2,\ldots.
\label{faf}
\end{equation}
With these operators one could work in a Hilbert space
with positive definite scalar product 
$(\Psi_1,\Psi_2)=\Psi^*_{1n}\Psi_{2n}$,
considering $a^+$ and $a^-$ as not basic
operators. 
However, due to concrete physical applications to be considered
in what follows, here we shall work in terms of operators $a^\pm$
using the corresponding indefinite scalar product.
The described possibility of employing hermitian conjugate
operators $f^+$ and $f^-$ will be discussed in last section.

\section{Guons, fermions and $q$-deformed Heisenberg algebra}
Let us suppose that $\nu\neq1$, and define the operators 
$
c^-=a^-G_\nu^{-1/2}(R),
$
$
c^+=G_\nu^{-1/2}(R) a^+,
$
$
G_\nu(R)=|1-\nu R|,
$
where for the moment we suppose that $R=(-1)^N$
with $N$ given by eq. (\ref{Nn}).
These operators anticommute with
reflection operator, $\{R,c^\pm\}=0$, and satisfy the
commutation relation
$
c^-c^+-G_\nu(R) G_{\nu}^{-1}(-R)c^+ c^-=sign(1+\nu R),
$
where $sign\, x$ is $+1$ for $x>0$ and $-1$ for $x<0$.
The operator $G_\nu(R)$ 
is reduced to
$G_\nu(R)=1-\nu R$ for
$-1<\nu<1$; for two other cases we have
$G_\nu(R)=\nu-R$,
$\nu>1$, and
$G_\nu(R)=R-(2p+1)$,
$\nu=-(2p+1)$. 
As a result, commutation relation is represented in first case
as
\begin{equation}
c^- c^+-g_\nu c^+c^-=1,
\quad
g_\nu=(1-\nu)^R(1+\nu)^{-R},\quad
-1<\nu<1,
\label{nu1a}
\end{equation}
whereas in two other cases it is reduced to 
\begin{equation}
c^- c^+-g_\nu c^+c^-=R,
\label{ccr}
\end{equation}
where $g_\nu=(\nu-1)^R(1+\nu)^{-R}$ for $\nu>1$ and
$g_\nu=p^R(1+p)^{-R}$ for $\nu=-(2p+1)$.
In the case corresponding to finite-dimensional representations
the final form (\ref{ccr}) has been obtained via additional
changing $R\rightarrow -R$. In all three cases operator-valued 
function $g_\nu$ satisfies the relation
$g_\nu c^\pm=c^\pm g^{-1}_\nu$. 
The deformed algebra of form (\ref{nu1a})
was introduced in ref. \cite{guon} in the context of generalized
statistics. The algebra (\ref{ccr}) represents some 
modification of (\ref{nu1a}).

The corresponding number operator $N=N(c^+,c^-)$ is given by 
\[
N=-\frac{\alpha}{2}+\frac{1}{2}\sqrt{|1-\nu^2|
(2c^+c^--\beta)
(2c^-c^+-\beta)+1},
\]
where 
$\alpha=-\nu$, $\beta=1$ in the case $-1<\nu<1$,
and $\alpha=-\nu+\nu^2-1$, $\beta=|\nu|$
for two other cases. 
Implying in relations (\ref{nu1a}), (\ref{ccr}) 
that  $R=(-1)^{N(c^+,c^-)}$, one 
can represent them in a closed form containing only
creation-annihilation operators $c^\pm$.
 
Let us take a limit $\nu\rightarrow \infty$ for the case $\nu>1$
and $p\rightarrow \infty$ for $\nu=-(2p+1)$
proceeding from relation (\ref{ccr}). 
Both cases lead to the algebra
\begin{equation}
c^-c^+-c^+ c^-=R,\quad \{R,c^\pm\}=0,\quad R^2=1.
\label{antir}
\end{equation}
Considering the Fock space representation
defined by relations $c^-|0\rangle=0$, $R|0\rangle=|0\rangle$, 
$\langle 0|0\rangle=1$,
one gets the relations $\langle 1|1\rangle=1$, $\langle 0|1\rangle=0$, 
and $\langle m|n\rangle=0$ for any $m\geq 2$ or $n\geq 2$,
where $|n\rangle=(c^+)^n|0\rangle$. It means that the (anti)commutation 
relations (\ref{antir}) 
have two-dimensional irreducible representation, in which 
$(c^\pm)^2=0$. In this case the operator $R$ is realized as
$R=1-2c^+c^-$, that reduces commutation
relations (\ref{antir}) 
to the standard fermionic anticommutation relations,
$
c^+c^-+c^+c^-=1,
$
$
c^{+2}=c^{-2}=0.
$
Therefore, fermionic algebra can be obtained from the guon-like form
of the $R$-deformed Heisenberg  algebra in the limit
$|\nu|\rightarrow+\infty$.

The substitution of operators $c^\pm$
into bosonic realization (\ref{abba}), (\ref{ab}) 
gives for $\nu\rightarrow \infty$
the well known realization of fermionic operators
in terms of bosonic operators $b^\pm$ \cite{ital,bos,bosany}:
\begin{equation}
c^-=\frac{\Pi_+}{\sqrt{N+1}}b^-,\quad
c^{^+}=(c^-)^\dagger.
\label{cbn}
\end{equation}
Here $\Pi_+$ and supplementary operator
$\Pi_-$,
$
\Pi_\pm=\frac{1}{2}(1\pm R),
$
are projector operators,
$
\Pi_\pm^2=\Pi_\pm,
$
$
\Pi_+\Pi_-=0,
$
$
\Pi_++\Pi_-=1,
$
and in eq. (\ref{cbn}) we mean that $R=(-1)^{N_b}$.
Due to relations $\Pi_\pm b^-=b^-\Pi_\mp$,
$\Pi_\pm b^+=b^+\Pi_\mp$, 
operators (\ref{cbn}) satisfy the standard
fermionic anticommutation relations.

This bosonization construction for fermions gives us a hint for
realization of $R$-paragrassmann algebra in terms of non-deformed
creation-annihilation operators $b^\pm$.
Indeed, the operators $a^\pm$ satisfying algebra
(\ref{aaa}), (\ref{parag}) can be realized as follows:
\begin{equation}
a^-=\varphi_p(N_b)F_p(N_b)b^-,\quad
a^+=b^+F_p(N_b)\varphi_p(N_b),
\label{abnb}
\end{equation}
where instead of projector operator 
$\Pi_+=\sin (\frac{\pi}{2}(N_b+1))$,
we have 
\begin{equation}
\varphi_p(N_b)=
\frac{\sin\left(\frac{\pi}{2p+1}(1+N_b)\right)}{\sin
\left(\frac{\pi}{2p+1}(1+{\cal N}_p)\right)}
\label{varn}
\end{equation}
with operator ${\cal N}_p={\cal N}_p(N_b)$,
\begin{equation}
{\cal N}_p(N_b)=p+\frac{1}{2}
-\left|N_b-\left(p+\frac{1}{2}+(2p+1)\left[\frac{N_b}{2p+1}\right]
\right)\right|.
\label{rhab}
\end{equation}
Here
$[X]$ is an integer part of $X$,
and operator $F_p(N_b)$ is given by
\[
F_p(N_b)=i^{N_b+1}\sqrt{\left|1-
\frac{2p+1}{2(N_b+1)}\left(1+(-1)^{N_b}\right)\right|}.
\]
Operator $\varphi_p$ has the properties
$
|\varphi_p(n)|=1,
$
$n\neq 2p\,\, mod(2p+1),$
$\varphi_p(2p+k(2p+1))=0,$
$k\in Z,$
and so,
$
\varphi_p(N_b)\varphi_p(N_b+1)\ldots\varphi_p(N_b+2p)=0.
$
Due to the latter property 
and relations
$G(N_b)b^\pm=b^\pm G(N_b\pm 1)$
being valid for any function $G(N_b)$,
one concludes that operators (\ref{abnb})
satisfy relations $(a^{\pm})^{2p+1}=0$.
They are odd operators, $Ra^\pm=-a^\pm R$, $R=(-1)^{N_b}$,
and satisfy $R$-deformed commutation relations (\ref{rdef}).
Since the relation
$(a^-)^\dagger\eta=\eta a^+$  takes place with $\eta=\eta^\dagger=
(-1)^{[(N_b+1)/2]}$,
the operators $a^-$ and $a^+$ are mutually
conjugate with respect to the indefinite scalar
product $(\Psi_1,\Psi_2)=$ $\langle\Psi_1|\eta\Psi_2\rangle$.
The obtained bosonized representation 
corresponds to finite-dimensional matrix
representation of the $R$-deformed Heisenberg algebra described
in the previous section.

Special form of fermionic algebra (\ref{antir})
can be generalized into the
algebra related to the $q$-deformed oscillator.
To realize such a generalization,
we note that since $R^2=1$,
$R$ is a phase operator. Then 
commutation relations
(\ref{antir}) ($R$-algebra) 
can be generalized into the $P$-algebra,
\begin{equation}
[a,\bar{a}]=P,
\label{palg}
\end{equation}
where $P$ is a phase operator with properties
generalizing the corresponding properties of operator 
$R$, 
\begin{equation}
P^p=1,\quad 
Pa=qaP,\quad
P\bar{a}=q^{-1}\bar{a}P,\quad
q=e^{-i\frac{2\pi}{p}},\quad
p=2,3,\ldots.
\label{pop}
\end{equation}
Using these relations, one finds that
the operators $a^p$ and  $\bar{a}{}^p$ commute with operators
$a$, $\bar{a}$ and $P$. In an irreducible representation
they are reduced to some constants. Assuming the existence
of the vacuum state $|0\rangle$, $a|0\rangle=0$, we find
that in Fock space representation of algebra (\ref{palg}), 
(\ref{pop}), there are the relations 
$
a^p=\bar{a}{}^p=0.
$
Multiplying relation (\ref{palg}) from the left 
by the operator $P^{-1}=P^{p-1}$,
we represent it in the form
of Lie-admissible algebra \cite{guon},
$
aT\bar{a}-\bar{a}Sa=1,
$
$
T=q^{-1}P^{-1},
$
$
S=qP^{-1}.
$
Defining new creation-annihilation operators,
$
c=q^{-1/2}aP^{-1/2},$
$\bar{c}=q^{-1/2}P^{-1/2}\bar{a},
$
one gets finally the $q$-deformed Heisenberg algebra
$
c\bar{c}-q\bar{c}c=1,
$
with deformation parameter $q$ being the primitive root of unity.

\section{OSp(2$|$2) supersymmetry}
The $R$-deformed Heisenberg algebra gives a possibility to
realize OSp(2$|$2) supersymmetry.
As a result we can get
unitary inifinite-dimensional half-bounded 
representations of $sl(2,R)$, $sl(2,R)\subset osp(1|2)\subset osp(2|2)$, 
and its non-unitary finite-dimensional representations.

In terms of generators of algebra (\ref{rdef}), the generators
of $osp(2|2)$ superalgebra can be realized as follows.
The even generators $J_0$, $J_\pm=J_1\pm iJ_2$ and 
$\Delta$ are given by relations
\begin{equation}
J_0=\frac{1}{4}\{a^-,a^+\},\quad
J_\pm=\frac{1}{2}(a^\pm)^2,\quad
\Delta=-\frac{1}{2}(R+\nu).
\label{jjj}
\end{equation}
They satisfy $sl(2,R)\times u(1)$ algebra,
\[
[J_0,J_\pm]=\pm J_\pm,
\quad 
[J_-,J_+]=2J_0,\quad
[\Delta,J_0]=[\Delta,J_\pm]=0.
\]
Odd generators $Q^\pm$, $S^\pm$ are represented as
$
Q^+=a^+\Pi_-,$
$Q^-=a^-\Pi_+,$
$S^+=a^+\Pi_+,$
$S^-=a^-\Pi_-,$
where $\Pi_\pm$ are the projectors, $\Pi_\pm=\frac{1}{2}(1\pm R)$. 
These operators satisfy the following anticommutation relations:
\begin{equation}
Q^{\pm2}=S^{\pm2}=0,\quad
\{S^+,Q^-\}=\{S^-,Q^+\}=0,
\label{qqss}
\end{equation}
\begin{equation}
\{Q^+,Q^-\}=2J_0+\Delta,\quad
\{S^+,S^-\}=2J_0-\Delta,\quad
\{S^+,Q^+\}=J_+,\quad
\{S^-,Q^-\}=J_-.
\label{qqh}
\end{equation}
Nontrivial commutators between 
even and odd generators are 
\begin{equation}
[J_+,Q^-]=-S^+,\quad
[J_+,S^-]=-Q^+,\quad
[J_-,Q^+]=S^-,\quad
[J_-,S^+]=Q^-,
\label{j+q}
\end{equation}
\begin{equation}
[J_0,Q^\pm]=\pm\frac{1}{2}Q^\pm,\quad
[J_0,S^\pm]=\pm\frac{1}{2}S^\pm,\quad
[\Delta,Q^\pm]=\pm Q^\pm,\quad
[\Delta,S^\pm]=\pm S^\pm.
\label{j0q}
\end{equation}
In the case $\nu>-1$,
the generators $J_0$, $J_\pm$ 
give the direct sum of half-bounded infinite-dimensional
unitary representations $D^+_{\alpha_+}$ and $D^+_{\alpha_-}$
of $sl(2,R)$,
being representations of the so called discrete series.
Here $\alpha_+=\frac{1}{4}(1+\nu)>0$,
and $\alpha_-=\alpha_++\frac{1}{2}$,
and these representations are realized on the subspaces spanned
by $|2n\rangle$ and $|2n+1\rangle$, $n=0,1,\ldots,$
where the corresponding Casimir operator of $sl(2,R)$,
$
C=-J_0^2+\frac{1}{2}\{J_+,J_-\},
$
takes the values $C=-\alpha_+(\alpha_+-1)$
and $C=-\alpha_-(\alpha_--1)$, 
and $J_0$ has the spectra $j_0=\alpha_++n$ and
$j_0=\alpha_-+n$, respectively \cite{bosany,any}.
In the case of the revealed finite-dimensional representations
of the $R$-deformed Heisenberg algebra, one finds
that the generators $J_0$, $J_\pm$
give a direct sum of two non-unitary 
$(p+1)-$ and $p-$dimensional irreducible
representations characterized by
the values of the Casimir operator
$C=-j_\pm(j_\pm+1)$ with
$j_+=p/2$ and $j_-=(p-1)/2$.
These representations are realized 
on the subspaces of even
and odd states, $|m\rangle_+= a^{+2m}|0\rangle$, $m=0,1\ldots,p$,
$|m\rangle_-=a^{+(2m+1)}|0\rangle$, $m=0,\ldots,p-1$,
where $J_0$ has the spectra
$j_0=(-j_+,-j_++1,\ldots,j_+)$ and $j_0=(-j_-,-j_-+1,\ldots,j_-)$. 
In other words, finite-dimensional representations
of the $R$-deformed Heisenberg algebra give finite-dimensional
representations of $(2+1)$-dimensional Lorentz group.
As it was noted in section 3,
the appearance of indefinite scalar 
product in the case of finite-dimensional
representations  means that such representations are non-unitary.
As we have seen, these representations
are the direct sum of finite-dimensional
representations of (2+1)-dimensional Lorentz group,
and since finite-dimensional representations of this group 
are non-unitary (see, e.g., ref. \cite{corply}),
we have here a `physical explanation' for non-unitarity of
finite-dimensional representations of the $R$-deformed
Heisenberg algebra with operators $a^+$ and $a^-$ to be mutually
conjugate.

In the simplest cases given by $p=1$ and $p=2$,
the corresponding metric operator in two-dimensional
even ($p=1$, $j_+=1/2$)
and odd ($p=2$, $j_-=1/2$) subspaces coincides 
up to a $c$-number factor with the operator $J_0$ being
restricted to the corresponding subspaces.
As a result, the indefinite scalar product on these
subspaces is the Dirac scalar product.
In the case of 3-dimensional vector
representations corresponding to $j_+=1$, $p=2$ and $j_-=1$, $p=3$,
the metric operator and generators $J_\mu$, $\mu=0,1,2,$ being
restricted to the corresponding even and odd subspaces, 
can be reduced by appropriate unitary transformation 
to the standard form of the 
vector realization with $(J_\mu)^{\nu}{}_\lambda=-i
\epsilon^{\nu}{}_{\mu\lambda}$ 
and $\eta_{\mu\nu}=diag (-,+,+)$ \cite{corply}.

As we have seen, the generators of $sl(2,R)$
algebra act reducibly in the cases
of infinite-dimensional and 
finite-dimensional representations of 
algebra (\ref{rdef}). On the other hand,
these generators together with operators
$a^\pm=Q^\pm+S^\pm$ give irreducible realization
of $osp(1|2)$ generators with
corresponding Casimir operator \cite{mac1,bem,bos,bosany}
$
{\cal C}=J_\mu J^\mu-\frac{1}{8}[a^-,a^+]$ 
taking the fixed value 
${\cal C}=\frac{1}{16}(1-\nu^2).$
 
In conclusion of this section we note that 
relations (\ref{qqss}) and (\ref{j0q})
mean that the pair of odd generators $Q^+$
and $Q^-$ together with even
generator $H_+=2J_0+\Delta$
form $s(2)$ superalgebra, 
$
Q^{\pm2}=0,
$
$
\{Q^+,Q^-\}=H_+,
$
$
[Q^\pm,H_+]=0,
$
whereas operators $S^+$ and $S^-$ are odd generators
of $s(2)$ superalgebra with even generator
$H_-=2J_0-\Delta$. 

\section{Outlook and concluding remarks}
The constructed guon-like algebra of the form
(\ref{nu1a}), (\ref{ccr}) contains
the operator-valued function $g_\nu$. But unlike
the original guon algebra \cite{guon},
here $g^2_\nu\neq 1$ and $[g_\nu,c^\pm]\neq 0$.
The condition of the form $[g,c^\pm]=0$ 
appeared in \cite{guon} from the requirement of micro causality 
under assumption that observables should be bilinear in 
fields or in creation-annihilation operators.
On the other hand, it is known that
in the field-theoretical anyonic constructions 
involving the Chern-Simons gauge field, there are
observables (e.g., total
angular momentum operator) which are not bilinear
in creation-annihiliation operators \cite{anycs}.
Moreover, the gauge-invariant fields carrying fractional
spin and statistics themselves turn out to be nonlocal operators
\cite{sred} being decomposable in some infinite series 
in degrees of creation-annihilation operators
of the initial matter field.
It seems that the guon-like algebra appeared here
could find some applications in the theory of anyons.

The revealed finite-dimensional representations of the $R$-deformed
Heisenberg algebra and their relationship to representations of
$(2+1)$-dimensional Lorentz group can be used for the construction
of universal minimal spinor set of linear differential equations
describing, on one hand,  
ordinary integer and half-integer spin fields and, on other hand,
fractional spin fields in $2+1$ dimensions.  Moreover, 
it is natural to try to apply these representations
for constructing $(2+1)$-dimensional supersymmetric field systems
since, as it was shown, any $(2p+1)$-dimensional
representation of the $R$-deformed Heisenberg algebra carries
the direct sum of spin-$j$,
$j=p/2$, and spin-$(j-1/2)$
representations of $(2+1)$-dimensional Lorentz group.
For the purpose, let us consider the simplest possible
nontrivial case corresponding to the choice of 5-dimensional
representation of the $R$-deformed Heisenberg algebra
with $\nu=-5$ ($p=2$), and construct the operators 
\[
{\cal D}_\alpha=\left(\frac{1}{2}-R\right){\cal P}_\alpha
-{\cal J}_\alpha
+\frac{1}{2}\epsilon m{\cal L}_\alpha,\quad
\epsilon=+,-.
\]
Here $\alpha=1,2$, $m$ is a mass parameter, 
${\cal P}_\alpha=-i(\gamma^\mu\partial_\mu)_\alpha{}^\beta 
{\cal L}_\beta$,
$\partial_\mu=\partial/\partial x^\mu$,
$x^\mu$ are external space-time coordinates independent from
$a^\pm$,
$\gamma_\mu$ is the set of (2+1)-dimensional $\gamma$-matrices
taken in the Majorana representation, 
$
(\gamma^{0})_{\alpha}{}^{\beta}=-(\sigma^{2})_{\alpha}{}^{\beta},
$ 
$(\gamma^{1})_{\alpha}{}^{\beta}=i(\sigma^{1})_{\alpha}{}^{\beta}$,
$(\gamma^{2})_{\alpha}{}^{\beta}=i(\sigma^{3})_{\alpha}{}^{\beta},$
${\cal L}_1=\frac{1}{\sqrt{2}}(a^++a^-)$,
${\cal L}_2=\frac{i}{\sqrt{2}}(a^+-a^-)$,
and 
${\cal J}_\alpha=
{\cal L}_\beta\epsilon_{\mu\nu\lambda}\partial^\mu J^\nu 
(\gamma^\lambda)_\alpha{}^\beta
$
with $J_\mu$ given by eq. (\ref{jjj}).
Operators ${\cal L}_\alpha$, ${\cal J}_\alpha$ and ${\cal P}_\alpha$
are spinor operators with respect to the action of the total
angular momentum vector operator, $M_\mu=
i\epsilon_{\mu\nu\lambda}x^\nu\partial^\lambda +J_\mu$,
$[M_\mu,M_\nu]=-i\epsilon_{\mu\nu\lambda}
M^\lambda$,
$[M_\mu,{\cal L}_\alpha]=
\frac{1}{2}(\gamma_\mu)_\alpha{}^\beta
{\cal L}_\beta$ etc., whereas the reflection operator $R$ is a scalar,
$[M_\mu, R]=0$. These properties of the operators
are, in fact, the consequence of the $osp(1|2)$ superalgebra generated 
by the operators $J_\mu$ and $a^\pm$, which has been discussed in the
previous section.
As a result, operator ${\cal D}_\alpha$ is
(2+1)-dimensional translation-invariant spinor 
operator. One can consider the set  of linear (in $\partial_\mu$)
differential field equations
\begin{equation}
{\cal D}_\alpha\Psi(x)=0
\label{dal}
\end{equation}
having in mind that $\Psi(x)$ is a 5-component field,
which with respect to (2+1)-dimensional
Lorentz group is transformed as
$\Psi(x)\rightarrow \Psi'(x')=\exp(iM_\mu\omega^\mu)\Psi(x)$,
where $\omega^\mu$ are the transformation parameters.
Therefore, eq. (\ref{dal})
is the covariant (spinor) set of (2+1)-dimensional field equations.
One can find that 
the field $\Psi(x)$ satisfying equations (\ref{dal})
is decomposed into the sum of fields $\Psi_\pm=\Pi_\pm \Psi$,
$\Psi=\Psi_++\Psi_-$,
carrying spins $s_+=-\epsilon$ and $s_-=\frac{1}{2}s_+$,
respectively. Field $\Psi_-$ is a 2-component
Dirac field, whereas 3-component field
$\Psi_+$ is, in fact, topologically massive 
Jackiw-Templeton-Deser-Schonfeld vector field \cite{vec,corply}.
Both fields have the same mass $m$, and, therefore,
spinor set of equations (\ref{dal}) describes a supermultiplet
of (2+1)-dimensional massive fields.  
The spinor supercharge operator generating the corresponding
supertransformations is
\begin{equation}
{\cal Q}_\alpha=\epsilon m{\cal L}_\alpha+
R{\cal P}_\alpha .
\label{qua}
\end{equation}
It anticommutes with ${\cal D}_\alpha$ on mass shell,
i.e. on the surface of equations (\ref{dal}),
$\{{\cal Q}_\alpha,{\cal D}_\beta\}\approx 0$,
and satisfies the relations $\{{\cal Q}_\alpha,
{\cal Q}_\beta\}\approx -16\epsilon 
m(\gamma_\mu\partial^\mu)_{\alpha\beta}$,
$[\partial_\mu,{\cal Q}_\alpha]=0$.
Now, one can introduce a field
$\Phi(x)$ carrying 
arbitrary (fixed) infinite- or finite-dimensional representation 
of the $R$-deformed Heisenberg algebra,
and consider another spinor set  of equations,
\begin{equation}
{\cal Q}_\alpha\Phi(x)=0.
\label{qal}
\end{equation}
Here operator ${\cal Q}_\alpha$ (\ref{qua})
is generalized to the case of the corresponding representation.
Solution of eq. (\ref{qal}) is decomposable into the
trivial field $\Psi_-(x)=\Pi_-\Phi(x)=0$ and 
field $\Phi_+(x)=\Pi_+\Phi(x)$ carrying irreducible representation
of the (2+1)-dimensional Poincar\'e group characterized by mass $m$
and spin $s_+=\epsilon\frac{1}{4}(1+\nu)$.
Therefore, spin of nontrivial field $\Phi_+$ is defined by the 
value of deformation parameter $\nu$, and one concludes 
that the spinor set
of equations (\ref{qal}) is the above-mentioned universal set
of linear differential equations giving some link between
fractional spin fields (anyons)
in the case of choosing $\nu>-1$ \cite{bosany,any},
and ordinary (2+1)-dimensional 
integer and half-integer spin fields in the case
$\nu=-(2p+1)$. 
The described (2+1)-dimensional supersymmetry as well as the
universal spinor set of linear differential field equations will be
considered in detail elsewhere \cite{prep}.
We only note here 
that the spinor sets of equations (\ref{dal}) and
(\ref{qal}) are analogous to (3+1)-dimensional Dirac
positive-energy equations
\cite{dir} in the sense that they represent by themselves
the covariant (spinor) sets of equations imposed on one
multi-component field.

In conclusion we note that in terms of hermitian conjugate operators
$f^+=a^+$ and $f^-=a^-R$ satisfying anticommutation relations
(\ref{faf}), the described finite-dimensional representations of
the $R$-deformed Heisenberg algebra 
supply us with some special
deformation of parafermionic algebra of order $2p$ with internal
$Z_2$ grading structure \cite{prep}.  Recently it was shown
\cite{deb} that new variants of parasupersymmetry can be
constructed with the help of finite-dimensional representations
of the $q$-deformed Heisenberg algebra.  It turns out that
physical properties of such new variants can be different from
the properties of the parasupersymmetry realized in terms of
the standard parafermionic generators
\cite{rubspi}.  Therefore, the revealed finite-dimensional
representations of the $R$-deformed Heisenberg algebra may 
also be interesting from the point of view of constructing
parasupersymmetric systems.  Perhaps, there the intrinsic
$Z_2$-grading structure of the corresponding deformed
parafermionic algebra could find physically interesting
consequences.

\vskip0.5cm
{\bf Acknowledgements}

The author thanks the Department of Physics
of Federal University of Juiz de Fora, where the part 
of this work was done, for hospitality.
The work was supported in part by RFFR grant No. 95-01-00249.

\end{document}